\documentclass[apj,twocolumn]{emulateapj}

\slugcomment{Accepted for publication in the ApJ, 2016 May 6}

\shortauthors{Parmentier}
\shorttitle{Star Formation Relations}

\newcommand\eff{\epsilon_{\rm ff}}
\newcommand\tff{\tau_{\rm ff}}
\newcommand\rhost{\rho_{\star}}

\newcommand\Sigst{\Sigma_{\star}}
\newcommand\Sigg{\Sigma_{\rm gas}}
\newcommand\Sigsfr{\Sigma_{\rm SFR}}

\newcommand\Ms{M_{\odot}}
\newcommand\Msy{M_{\odot} \cdot yr^{-1}}
\newcommand\Mspp{M_{\odot} \cdot pc^{-2}}
\newcommand\Msppp{M_{\odot} \cdot pc^{-3}}

\newcommand\fft{free-fall time }
\newcommand\sfe{star formation efficiency }
\newcommand\sfr{star formation rate }
\newcommand\stf{star formation }
\newcommand\sfing {star-forming }

\newcommand\SoN{Solar neighborhood }

\begin{document}



\title{A cautionary note about composite Galactic star formation relations}


\author{G.~Parmentier\altaffilmark{1}}


\altaffiltext{1}{Astronomisches Rechen-Institut, Zentrum f\"ur Astronomie der Universit\"at Heidelberg, M\"onchhofstr. 12-14, D-69120 Heidelberg, Germany}


\begin{abstract}
We explore the pitfalls which affect the comparison of the \stf relation for nearby molecular clouds with that for distant compact molecular clumps.  We show that both relations behave differently in the ($\Sigg$, $\Sigsfr$) space, where  $\Sigg$ and $\Sigsfr$ are, respectively, the gas and \stf rate surface densities, even when the physics of star formation is the same.  This is because the \stf relation of nearby clouds relates gas and star surface densities measured locally, that is, within a given interval of gas surface density, or at a given protostar location.  We refer to such measurements as local measurements, and the corresponding \stf relation as the local relation.  
In contrast, the stellar content of a distant molecular clump remains unresolved.  Only the mean \stf rate can be obtained from e.g. the clump infrared luminosity.  One clump therefore provides one single point to the ($\Sigg$, $\Sigsfr$) space, that is, its mean gas surface density and \stf rate surface density.  We refer to this \stf relation as a global relation since it builds on the global properties of molecular clumps.  Its definition therefore requires an ensemble of cluster-forming clumps.  We show that, although the local and global relations have different slopes, this per se cannot be taken as evidence for a change in the physics of \stf with  gas surface density.  
It therefore appears that great caution should be taken when physically interpreting a composite \stf relation, that is, a relation combining together local and global relations.
\end{abstract}



\section{Introduction}
\label{sec:intro}

The \sfr of a galaxy is a crucial driver of its evolution \citep{fri94}.  Its dependence on the gas density and/or galaxy type is often embodied by \stf relations, namely, power-law relations between the surface density of the gas, $\Sigg$, and the surface density of the star formation rate, $\Sigsfr$:

\begin{equation}
\Sigsfr \propto \Sigg^N\;.
\label{eq:sfla}
\end{equation}

Based on a combined sample of spiral and starburst galaxies, \citet{ken98a} derive a super-linear relation ($N \simeq 1.4$; his fig.~6) between the disk-averaged surface densities of the \sfr ($\Sigsfr$) and of the total gas mass ($\Sigg$), where the latter accounts for the atomic and molecular hydrogen (obtained via H~I and CO measurements, respectively).  Equation~\ref{eq:sfla} with $N \simeq 1.4$ is often referred to as the Schmidt - Kennicutt law.  \citet{sch59} was the first to postulate a power-law scaling for the neutral hydrogen and Population~I stars of the Galactic disc, for which he found $N \simeq 2$.  He also suggested that Eq.~1 could be valid for galaxies. 

In addition to surface-density measurements, \stf relations can also be defined based on global measurements, i.e. as the total \stf rate as a function of gas mass.  In that respect, \citet{gs04} investigate how the \stf activity of a sample of spiral, luminous and ultraluminous infrared galaxies scales with their total and dense molecular gas contents.  They find a super-linear (slope $\simeq 1.4$) relation between the total \sfr (as traced by the galaxy far-infrared luminosity, $L_{IR}$) and the total molecular gas mass (as traced by CO luminosity, $L_{CO}$), i.e. $L_{IR} \propto L_{CO}^{1.4}$.  That is, the slope is reminiscent of that of \citet{ken98a}.  In contrast, the relation between the \sfr of galaxies and their dense molecular gas mass (as traced by the HCN line luminosity) is linear, i.e. $L_{IR} \propto L_{HCN}$.  This direct proportionality between the \sfr and the dense gas mass thus differs from the Schmidt - Kennicutt relation.
According to \citet{gs04}, the linear scaling $L_{IR} \propto L_{HCN}$ stems from the high-mass star-forming sites being the dense regions of giant molecular clouds, rather than their envelopes which are the main CO-emitters.  This is the combination of spiral and starburst galaxies in one single sample which yields $N \simeq 1.4$ in the Schmidt - Kennicutt relation and in the $L_{\rm CO} - L_{\rm IR}$ relation of \citet{gs04}.  In starburst, luminous and ultraluminous infrared galaxies, the dense gas mass fraction is higher than in spiral galaxies, thereby promoting a higher \stf activity for a given $L_{\rm CO}$ luminosity.  

This interpretation has been strengthened by \citet{wu05} who find that the linear correlation between $L_{\rm HCN}$ and $L_{\rm IR}$ established by \citet{gs04} for entire galaxies also holds for individual molecular clumps of the Galactic disk.  Following \citet{gs04}, \citet{wu05} therefore also argue that the most relevant parameter for the \sfr is the amount of {\it dense} molecular gas traced by HCN emission, that is, gas with a mean density of $\simeq 3 \cdot 10^4 cm^{-3}$.  Further support to this picture has been added by \citet{lad12} who find that the total star formation rate of nearby molecular clouds is linearly proportional to their dense gas content \citep[see also][]{lad10}, and that this relation for molecular clouds might extend that already established for galaxies.  
In fact, regardless of whether one considers their total gas mass or their dense gas mass, nearby clouds do not show any super-linear \stf relation.  That is, the data for the nearby clouds \citep{lad12} and for spiral galaxies \citep{gs04} are incompatible with the classical Schmidt - Kennicutt relation.  \citet{lad12} show that the \stf relation is linear, from individual clouds to entire galaxies, as long as their dense gas fraction is the same. 

In addition to the global measurements for galaxies and molecular clouds quoted above, \stf relations have also been established for the interiors of molecular clouds of the \SoN \citep[e.g.,][]{hei10, gut11, lom13, lad13}:
\begin{equation}
\Sigst \propto \Sigg^{N'} \,,
\label{eq:sflb}
\end{equation}

where $\Sigg$ and $\Sigst$ are the surface densities of gas and young stellar objects (YSOs) measured in a particular region of a molecular cloud, rather than averaged over the entire cloud.  The spatial scale of these \stf relations is therefore the pc-scale, and the measured slope ranges from $N' \simeq 2$ \citep{gut11, lom13, lad13} to $N'\simeq 3$ \citep{hei10, lad13}.  By assigning a duration to the \stf episode, $t_{SF}$, one obtains the surface density of the \stf rate: 
\begin{equation}
\Sigsfr = \frac{ \Sigst }{t_{SF}} \propto \Sigg^{N'} \,,
\label{eq:sflc}
\end{equation}
an equation analogous to Eq.~\ref{eq:sfla} although, given the local nature of its measurements, Eq.~\ref{eq:sflc} cannot be directly compared to Eq.~\ref{eq:sfla}.

Nearby molecular clouds are exquisite targets when it comes to resolving their stellar content on a YSO-by-YSO basis, but they are devoid of massive stars.  To learn about massive-star formation, we must look at compact molecular clumps with distances of, typically, several kpc.  At such distances, their embedded stellar content remains unresolved (although ALMA is now changing that).  Their \stf activity must thus be traced by a global property such as the clump total infrared luminosity, $24\,\mu m$ emission, or radio continuum emission \citep{vut13}.   

\citet{wu10} obtain the total ($8 - 1000\,\mu m$) infrared luminosity, $L_{IR}$, and the mean gas surface density of about 50 molecular clumps of the Galactic disk.  In a comprehensive study aimed at comparing the \stf relations of nearby molecular clouds, distant molecular clumps and galaxies, \citet{hei10} compare the gas and \stf rate surface densities of these clumps to the \stf relation they have derived for molecular clouds of the Solar neighborhood.  To obtain the \stf rate surface density of the clumps, they assume the conversion between the total infrared luminosity and \sfr established for galaxies by \citet{ken98b}.  

They find that the \stf relation for molecular clumps is shallower than that for molecular clouds \citep[see Figure 9 in ][]{hei10}.  Also, at gas surface densities of $\Sigg \simeq 300\,\Mspp$, an increase of the clump \sfr by a factor of 3-to-10 is needed to bring it in agreement with the values obtained for nearby molecular clouds.  \citet{hei10} quote two possibilities to explain this difference: 
(i)~star-forming regions behave differently in the ($\Sigg$, $\Sigsfr$)-space depending on whether they form massive stars (as for distant clumps), or not (as for nearby molecular clouds); (ii)~the $L_{IR} - SFR$ calibration of \citet{ken98b} requires an upward correction.    
   
In this contribution, we discuss a third reason as to why there is a shift between the \stf relations for distant clumps and nearby clouds.  This effect is the following.  While the gas and \sfr surface densities of distant clumps are averaged over the entire clump extent, in nearby clouds, they can be measured {\it at a given location} of the cloud.  In what follows we will refer these measurements as global and local, respectively.      
A local meaurement in a nearby cloud can be done at a given extinction level \citep[e.g.][]{hei10}, or at a given YSO location \citep[e.g.][]{gut11}.  In that respect, we stress that the term `local' does not refer to the distance of the clouds (although due to the resolving power needed to identify individual YSOs, local measurements have so far been performed in nearby molecular clouds only).  We will show that measuring the gas and star surface densities globally or locally inevitably leads to different \stf relations, in terms of both normalization and slope, {\it even when the underlying physics of star formation is the same}.  Displaying together the local and global \stf relations yields a broken power-law whose composite nature hinders the physical interpretation.  In particular, the steepening of the composite slope at low gas surface densities does not constitute an evidence of a star-formation threshold, at least not a threshold defined in a Heaviside manner in the ($\Sigg$, $\Sigsfr$)-space.  That local and global measurements lead to distinct \stf relations have already been highlighted by \citet{lad13} who show that there is a difference  between the local and global \stf relations of nearby molecular clouds.  That is, there is a Schmidt relation within molecular clouds, but not between molecular clouds given that Galactic clouds have a similar averaged surface density.     

The paper is organized as follows.  In Section \ref{sec:princ}, we show how global and local \stf relations differ from each other.  We also show how the mass-radius relation of molecular clumps drives the slope of their global \stf relation.  Section~\ref{sec:steep} compares models and observations of the local relation.  In Section~\ref{sec:break}, we emphasize that the difference in slope between the local and global relations {\it per se} does not indicate a change in the physics of star formation with gas surface density.  We also discuss the implications in terms of \stf density threshold.
In Section~\ref{sec:sfeff}, we discuss the difficulties inherent to estimating the \sfe per \fft from the global and local relations.
Finally, we present our conclusions in Section~\ref{sec:conclu}.

\section{Global vs. Local Star Formation relations}
\label{sec:princ}

Molecular clumps are individual regions of massive-star and star-cluster formation.  Their size is often quantified with the full-width at half-maximum (FWHM) of a given molecular tracer.  As an example, \citet{wu10} map molecular clumps in the ${\rm HCN} \,\, J = 1-0$ transition, and define their radius, $R_{HCN}$, as half the FWHM of the corresponding radial intensity profile, i.e., $R_{HCN} = 0.5\,FWHM_{\rm HCN}$.  The gas surface density, $\Sigg$, is then averaged over a disk of radius $R_{HCN}$:

\begin{equation}
\Sigg  = \frac{M_{gas}}{\pi R_{HCN}^2}\,,
\end{equation}    
with $M_{gas}$ the gas mass enclosed within the half-peak intensity contour \citep[see Table~11 in][]{wu10}.  

To estimate the \stf rate, $SFR$, of a clump, \citet{hei10} use the clump total infrared luminosity, $L_{IR}$ (8-1000\,$\mu$m) \citep[see Table 7 in][]{wu10}, and assume the conversion of \citet{ken98b}:

\begin{equation}
SFR_{IR} (\Msy) \simeq 2 \times 10^{-10} L_{IR} (L_{\odot})\;.
\label{eq:ken98}
\end{equation} 

Note that this relation was introduced for optically-thick starburst galaxies, whose bolometric stellar luminosity is reprocessed in the infrared.  We will further discuss Eq.~\ref{eq:ken98} in section~\ref{ssec:LSFR}.    

The clump \sfr surface density, $\Sigma_{\rm SFR_{IR}}$, is also averaged over a disk of radius $R_{HCN}$:

\begin{equation}
\Sigma_{\rm SFR_{IR}} = \frac{\rm SFR_{IR}}{\pi R_{HCN}^2}
\end{equation}    
\citep[see Eq.~16 and Table~6 in][]{hei10}.  

We refer to $\Sigg$ and $\Sigma_{\rm SFR_{IR}}$ as {\it global} surface densities since they are averaged over the whole FWHM-extent of the clump.

For nearby \sfing regions, however, it is possible to measure the surface densities of gas and YSOs in successive intervals of visual extinction nested inside a given molecular cloud.  That is, clouds are divided into contour levels of gas surface density, and the gas mass, YSO number and surface area are measured for each interval \citep[see Fig.~2 in ][]{hei10}.  We refer the corresponding surface densities as {\it local} surface densities.  

\begin{figure}
\begin{center}
\epsscale{1.0}  \plotone{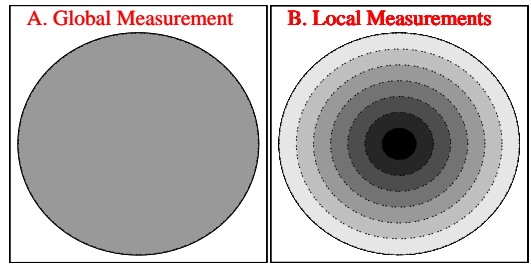}
\caption{Illustration of global and local measures of surface densities.  In a global measure (left panel), the surface densities of the gas and stars (or of the star formation rate) are averaged over the whole molecular clump.  In a local measure (right panel), the surface densities are averaged within given contour intervals of the gas surface density.  
Whether the surface densities are measured locally or globally leads to different \stf relations (see Section \ref{sec:princ}). }
\label{fig:sk}
\end{center} 
\end{figure}

The difference in how both measurements are carried out is illustrated in Fig.~\ref{fig:sk} for the case of a spherical reservoir of gas.  It should be noted that in case of global measurements (left panel of Fig.~\ref{fig:sk}), to define a \stf relation requires a {\it population} of molecular clumps (since each clump provides one point in the ($\Sigg$, $\Sigsfr$)-space).  In case of local measurements (right panel of Fig.~\ref{fig:sk}), a molecular cloud or, even, one only of its embedded clusters, is enough to define a \stf relation.  In what follows, we will refer to the \stf relation of \citet{hei10} as a local relation, and the \stf relation of an ensemble of dense clumps as its global counterpart.       

We note that the \stf relations obtained by \citet{gut11} for a sample of nearby molecular clouds is also a local relation since the gas and YSO surface densities are measured in the vicinity of each YSO based on a nearest-neighbor scheme.       
 
In the next section, we build on the concept of \sfe per \fft to obtain the global star formation relation of molecular clumps as a function of their mass-radius relation.  We will assume that the \sfe per \fft is a constant.   This seems to be a reasonable hypothesis on the spatial scale of a forming-cluster (i.e. a few pc) since the slope of the local relation predicted by the model of \citet{par13} ($N' \simeq 2$) matches the observed one \citep{gut11, lom13, lad13} when the \sfe per \fft is constant.  If it were to increase with gas density, the local \stf relation would be steeper, a possibility we discuss in Section~\ref{sec:steep}.

\subsection{The Global Star Formation Relation}
\label{ssec:glob}
Consider a molecular clump with a gas mass $M_{g}$ enclosed within a radius $R$.  Here, we remind that molecular clumps lack a neat outer boundary.  Their radius is often defined as half the FWHM of a given gas-tracer, with the clump mass the gas mass enclosed inside the half-peak intensity contour.  The mean surface densities of the gas and \sfr of the clump are, respectively:
\begin{equation}
\Sigg^{glob} (t) = \frac{ M_{g}(t) }{\pi R^2 }
\label{eq:siggg}
\end{equation}
and 
\begin{equation}
\Sigsfr^{glob} (t)= \frac{{\rm SFR}(t)}{\pi R^2 }\,,
\label{eq:sigstg}
\end{equation}
where $t$ is the time elapsed since the onset of star formation.  We assume that a clump is an isolated system (i.e. no inflows or outflows) and that it is in static equilibrium.  As a result, the surface densities of the gas and \sfr decrease with time as a result of gas-feeding to \stf \citep[see ][for a discussion about the evolution of the star formation rate]{par14}. 

The global \sfr of the clump obeys: 
\begin{equation}
SFR(t) \simeq \eff \frac{M_g(t)}{\tff(t)}\,,
\label{eq:sfr}
\end{equation}
with $\eff$ the \sfe per \fft and $M_g(t)$ the gas mass not yet processed into stars at time $t$.  $\tff$ is the free-fall time at the mean volume density of the clump gas, i.e.:
\begin{equation}
\tff(t) = \sqrt{ \frac{3 \pi}{32 G \rho_g(t)} }\,,
\end{equation}
with $\rho_g (t)= 3 M_g(t) / (4\pi R^3)$ and $G$ the gravitational constant.  
As we shall see later, Equation~(\ref{eq:sfr}) slightly underestimates the actual \sfr of centrally-concentrated molecular clumps.  
 
The surface density of the clump \sfr is then simply: 
\begin{equation}
\Sigma_{SFR}^{glob}(t) = \frac{SFR(t)}{\pi R^2} = \frac{\eff}{\tff(t)} \cdot \frac{M_g(t)}{\pi R^2} = \frac{\eff}{\tff(t)} \cdot \Sigg^{glob} (t) \;.
\label{eq:sigsfr}
\end{equation}
To define a \stf relation based on Eq.~\ref{eq:sigsfr}, one needs to consider an {\it ensemble} of molecular clumps.  Let us first consider clumps with a common mean volume density and, therefore, the same  free-fall time.  Equation \ref{eq:sigsfr} shows that, with $\eff$/$\tff$ a constant among the clump population, the slope of the global \stf relation is unity ($N=1$).  Its normalization depends on the clump volume density and on the \sfe per free-fall time.  The denser the clump and/or the higher the \sfe per free-fall time, the higher the \sfr surface density.  [We will look into the time evolution later in this section]

Now let us consider a population of clumps with identical radii, rather than identical volume densities.  Equation~(\ref{eq:sigsfr}) can be rewritten as:  
\begin{equation}
\Sigma_{SFR}^{glob}(t) = 2 \eff \sqrt{\frac{ 2G }{\pi R}} \cdot [\Sigma_g(t)]^{3/2}\;.
\label{eq:sigsfr_R}
\end{equation}
The \stf relation is now steeper ($N=1.5$) than found for a common mean volume density ($N=1$).  That is, the slope of the global \stf relation depends on the assumed clump mass-radius relation.  At constant radius, the volume density of the gas increases along with its surface density, which speeds up star formation as $\Sigg$ increases.

If the star-forming molecular clumps are characterized by a common mean surface density, then the \stf relation becomes a vertical in the ($\Sigg$, $\Sigsfr$) space.  The same pattern holds for giant molecular clouds, given their common mean surface density, as already highlighted by \citet{lad13} (see their fig.~8).  We therefore note that to define a global \stf relation for molecular clumps is an ambiguous task, as the slope depends on the clump mass-radius relation.    

\begin{figure}
\begin{center}
\epsscale{1.1}  \plotone{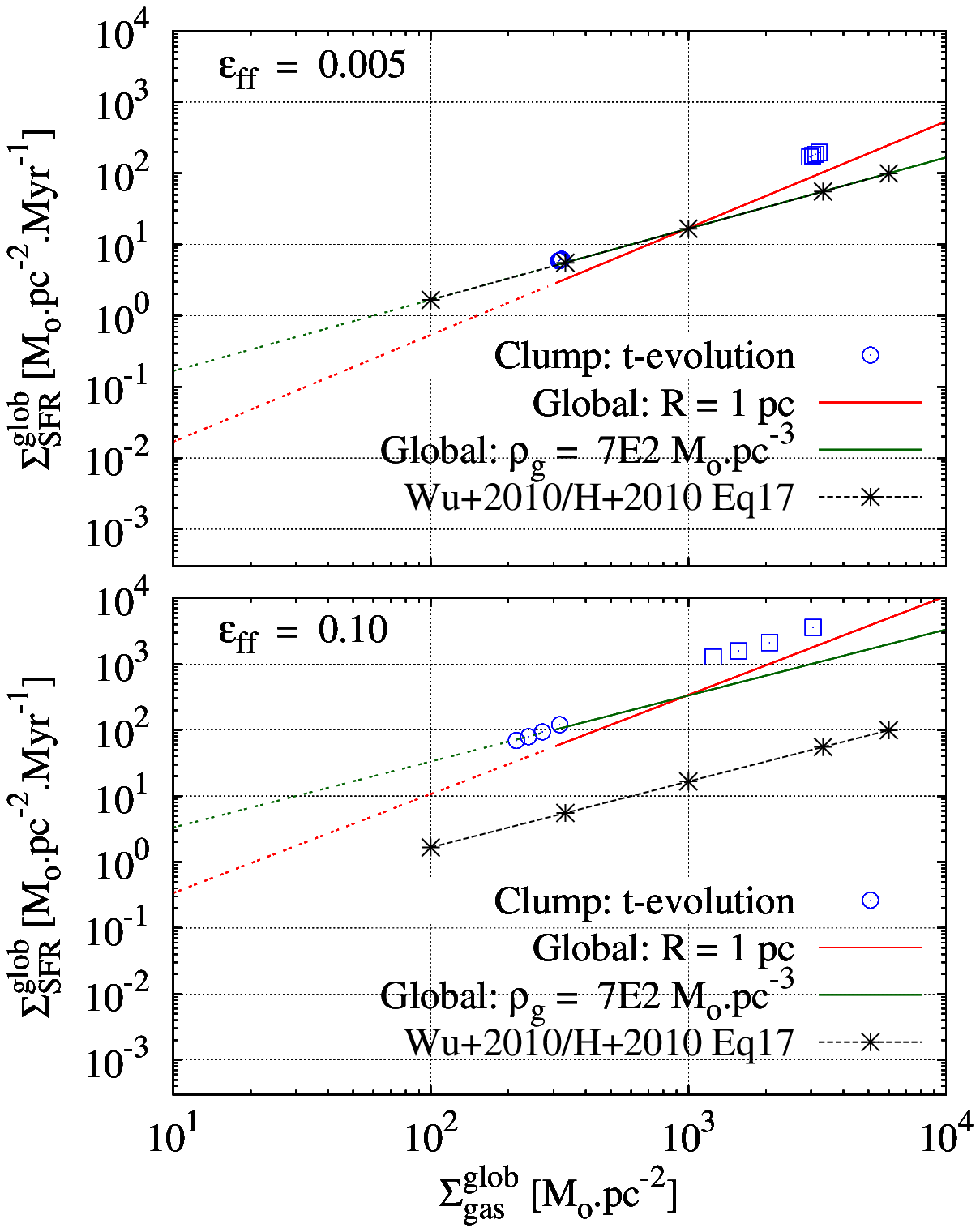}
\caption{Global \stf relations for two populations of molecular clumps, one with a given mean volume density of 700$\Ms\cdot pc^{-3}$ (Eq.~\ref{eq:sigsfr}, green line) and one with a given radius of 1\,pc (Eq.~\ref{eq:sigsfr_R}, red line).  The black line with asterisks is Equation~17 of \citet{hei10}, which shows the region of the diagram where the clumps mapped in HCN\,1-0 by \citet{wu10} are located.  The blue open symbols show how the surface densities evolve with time for clumps with a radius of 1\,pc, and masses of $10^3\Ms$ (circles) and $10^4\Ms$ (squares).  The time-evolution is from right to left and the displayed time-span is 2\,Myr.  The \sfe per \fft is $\eff = 0.005$ and $\eff = 0.10$ in the top and bottom panels, respectively (see Section~\ref{ssec:glob} for details).  }
\label{fig:glob}
\end{center} 
\end{figure}

Equations~(\ref{eq:sigsfr}) and (\ref{eq:sigsfr_R}) are shown in both panels of Fig.~\ref{fig:glob} as the green and red lines, respectively.  Equation~(\ref{eq:sigsfr}) uses a \fft of 0.3\,Myr, which corresponds to a mean volume density of $700\,\Msppp$, or a number density of $n_{H2} \simeq 10^4\,cm^{-3}$.  This is the median number density of the clumps mapped in HCN~1-0 by \citet{wu10}.  An order of magnitude for the radius of these clumps is 1\,pc, and we therefore adopt $R=1\,pc$ in Equation~(\ref{eq:sigsfr_R}).  

In the top panel of Fig.~\ref{fig:glob}, a \sfe per \fft of $\eff = 0.005$ is adopted so that Equation (\ref{eq:sigsfr}) (the green line) matches Equation~17 of \citet{hei10} (black line with asterisks).  That equation, which we reproduce here for the sake of clarity,
\begin{equation}
\Sigsfr \simeq 1.66 \times 10^{-2} \left( \frac{\Sigma_{{\rm HCN}(1-0)}}{1 \Mspp} \right) (\Ms \cdot Myr^{-1} \cdot pc^{-2})\,,
\label{eq:hei10}
\end{equation}
provides a good match to the locus of the HCN 1-0 clumps of \citet{wu10} (see also the middle panel of Fig.~\ref{fig:hei10wu05} where Eq.~17 of \citet{hei10} is shown along with the data of \citet{wu10}). 

In the bottom panel of Fig.~~\ref{fig:glob}, the \sfe per \fft is $\eff = 0.10$, which increases the intercept of Eqs~\ref{eq:sigsfr} and \ref{eq:sigsfr_R} by a factor of 20 compared to the top panel.  $\eff = 0.10$ is the \sfe per \fft used by \citet{par13} to match their model onto the observed local \stf relation of \citet{gut11}.

It is puzzling that there is such a difference in the \sfe per \fft needed to match the global surface densities of distant dense clumps on the one hand (top panel), and the local surface densities of nearby clouds on the other \citep[see fig.~3 in ][]{par13}.  The discrepancy may stem from either observed clump stellar surface densities being underestimated, or observed cloud stellar surface densities being overestimated, or a combination of both.  We will further discuss this issue in  Section~\ref{sec:sfeff}.

The blue open circles and squares indicate how an isolated clump in static equilibrium evolves with time in the ($\Sigg^{glob}$, $\Sigsfr^{glob}$) parameter space according to the semi-analytical model of \citet{par13}.  In Fig.~\ref{fig:glob}, the \sfr of these model clumps is averaged over time.  That is, it is defined as the total stellar mass built at time $t$, $M_{\star}^{glob}(t)$, divided by $t$.  The \sfr surface density thus obeys: 
\begin{equation}
\Sigma_{SFR}^{glob}(t) = \frac{ M_{\star}^{glob} (t)}{\pi R^2 \cdot t} = \frac{ \Sigst^{glob} (t)}{t}\;.
\label{eq:sigcode}
\end{equation}

As part of the gas is used for star formation, the \sfe per \fft is applied to an ever lower gas mass over an ever longer free-fall time.  As a result, the clump location moves toward smaller surface densities of both the gas and \stf rate.  The adopted clump radius is 1\,pc, the initial gas masses are $10^3\Ms$ (circles) and $10^4\Ms$ (squares), and the physical time-span is 2\,Myr. 

As expected, the amount of evolution is larger for higher $\eff$ and/or higher densities.  That is, the span in surface densities is larger in the bottom panel ($\eff = 0.10$) than in the top panel ($\eff = 0.005$), and is also larger for the squares than for the circles.  
The length of the sequence also depends on the adopted time-span.  Here it is assumed that cluster-forming clumps retain their gas during at least 2\,Myr.  This time-span remains uncertain, however.  The time-scale needed for an embedded cluster to expel its residual \sfing gas depends on an array of parameters, ranging from the depth of the forming-cluster gravitational potential \citep[hence the clump mass and radius; ][]{par08}, to the kinetic energy released into the intra-cluster medium by the newly formed massive stars \citep{dib11}.  The kinetic energy itself depends on the clump mass, \stf efficiency, the stellar initial mass function and how-well it is sampled in the high-mass regime.  Another factor at work is the duration of the pre-main sequence phase as a function of stellar mass.  Star-forming regions of the \SoN contains Class~II and Class~III YSOs.  Given that the life-time of a Class~II object is $\simeq 2$\,Myr \citep{eva09}, this means that these regions have been experiencing \stf for at least 2\,Myr \citep[but see ][for a longer estimate]{bell13}.  Dense and massive molecular clumps, however, contain intermediate-mass and massive pre-main-sequence stars which settle on the main sequence within significantly less than 2\,Myr.  Stellar winds, and stellar feedback in general, may thus terminate the cluster embedded stage at an  earlier time than in the Solar neighborhood.  Our adopted time-span of 2\,Myr may thus be an upper limit in the case of dense and massive clumps, as may be the length of the sequence of blue open symbols in Fig.~\ref{fig:glob}.

Given that the model clump radius is 1\,pc, one would expect to find the starting point of each blue sequence on the red line, which depicts the global \stf relation for a population of clumps with $R=$1\,pc (Equation \ref{eq:sigsfr_R}).  The shift between the red lines and the blue-sequence starting points arises because the latter accounts for the concentrated nature of molecular clumps.  
Star formation proceeds faster in clumps with volume-density gradients than in clumps with a flat density profile.  
The initial gas volume density profile adopted for the open blue symbols of Fig.~\ref{fig:glob} is an isothermal sphere (i.e. $\rho_{gas}(r, t=0) \propto r^{-2}$ with $r$ the distance to the clump center).  
For such a density profile, the global \sfe achieved within one free-fall time is not $\eff$, as may be naively expected, but $1.6\eff$ \citep[see][]{par14a}.  The clump density profile cannot be taken into account by the simple Eq.~\ref{eq:sfr}, from which Equation~\ref{eq:sigsfr_R} is derived.  Shifting upwards the analytical solution given by Equation~\ref{eq:sigsfr_R} (red line) by a factor 1.6 brings it in agreement with the numerical solution illustrated by the blue open symbols.     

A point worth being emphasized in this section is that even if star formation operates on a given \sfe per free-fall time, the slope of the \stf relation is {\it not necessarily} $N=1.5$.    
The Kennicutt-Schmidt relation (for galaxies) has an index $N=1.4$ \citep{ken98a}.  This is often taken as an evidence that star formation proceeds with a given \sfe per \fft since the \sfr then scales as the gas mass divided by the gas free-fall time (Eq.~\ref{eq:sfr}).  Expressed in terms of the volume densities of gas, $\rho_{gas}$, and \stf rate, $\rho_{SFR}$, this gives: 
\begin{equation}
\rho _{SFR} \propto \frac{\rho_{gas}}{(G \rho_{gas})^{-0.5}} \propto (\rho_{gas})^{1.5}\,,
\end{equation}

and, if the gas scale-height does not vary strongly from galaxy-to-galaxy: 

\begin{equation}
\Sigma _{SFR} \propto (\Sigma_{gas})^{1.5}\,,
\end{equation}

in agreement with the observed Kennicutt-Schmidt relation \citep[see e.g.][]{won02, kru07b}.

This line-of-reasoning, however, implies that the \stf relation spans a range of volume densities.  As we have seen above, when considering star-forming units with a limited range of mean volume densities (e.g. clumps mapped in HCN $J = 1 - 0$), the \stf relation is as given by Eq.~\ref{eq:sigsfr} with $\tff$ roughly constant and, thus, $N=1$.  Therefore, an observed global \stf relation with $N \neq 1.5$ does not necessarily disprove the hypothesis that \stf proceeds with a given \sfe per free-fall time.

\subsection{The Local Star Formation Relation}
\label{ssec:loc}

Having dealt with the global properties of clumps, we next investigate how the stars and gas are distributed {\it inside} them.  
That is, for a given clump, we now focus on the radial profiles of the gas and star surface densities, $\Sigg^{loc} (s,t)$ and $\Sigst^{loc} (s,t)$, and we obtain the corresponding \stf relation. $t$ is the time elapsed since the onset of star formation, and $s$ is the  projected distance from the clump center (while we denote $r$ its three-dimensional counterpart).  The clump can be a compact high-density one, or a larger and more diffuse one (e.g. a concentration of embedded YSOs in a nearby molecular cloud).  Spherical symmetry is assumed in any case.  

To obtain $\Sigst^{loc} (s,t)$ and $\Sigg^{loc} (s,t)$, we build on the model of \citet{par13} which predicts the \stf history of isolated spheres of gas in static equilibrium.  The principle of this model is to associate a given \sfe per free-fall time, $\eff$, to an initial gas density profile, $\rho_0(r)$.  The volume density profile of the gas not yet turned into stars at time $t$, $\rho_g(t,r)$, and the volume density profile of the forming cluster, $\rhost(t,r)$, are given by Equations~(19) and (20) in \citet{par13}, respectively.  We reproduce them below for the sake of clarity:
\begin{equation}
\rho_g(r,t) = \left( \rho_0(r)^{-1/2} + \sqrt{\frac{8G}{3\pi}} \cdot \eff \cdot t   \right)^{-2}\,,
\label{eq:rhog}
\end{equation}
and
\begin{equation}
\rhost(r,t) = \rho_0(r) - \left( \rho_0(r)^{-1/2} + \sqrt{\frac{8G}{3\pi}} \cdot \eff \cdot t   \right)^{-2}\;.
\label{eq:rhost}
\end{equation}
We emphasize that the model is purely volumetric, i.e. the \sfe per \fft is applied to the volume density, with the conversion to surface density being performed for the sole purpose of comparing the model to  observations.  The projection in two dimensions of Equations~(\ref{eq:rhog})-(\ref{eq:rhost}) provides the relation between $\Sigst^{loc} (s,t)$ and $\Sigg^{loc} (s,t)$ \citep[see Fig.~3 in][]{par13}.  This is the {\it local} star formation  relation, that is, the \stf relation defined {\it all through} a clump, as opposed to the {\it global} star formation relation defined for a {\it population} of clumps.

\begin{figure}
\begin{center}
\epsscale{1.1}  \plotone{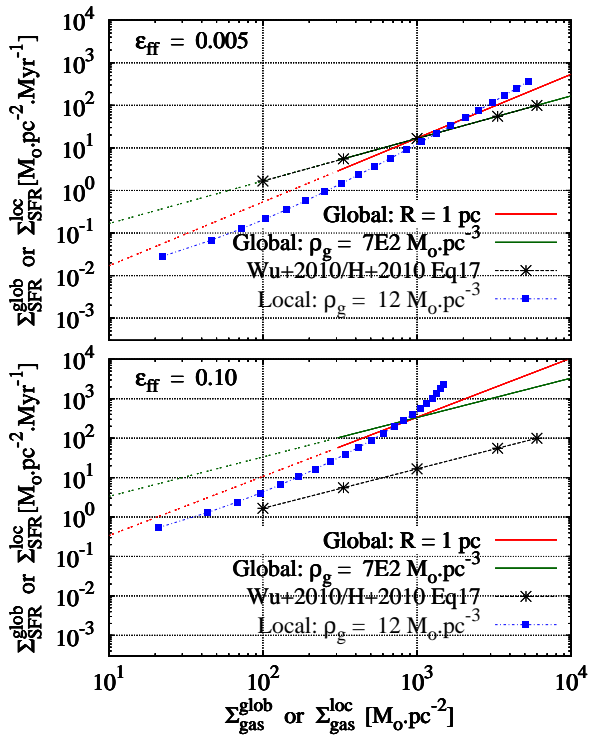}
\caption{Same as Figure \ref{fig:glob}, but completed with the local \stf relation for a molecular clump of radius 6\,pc and total mass $10^4\,\Ms$ (blue line with plain squares).}
\label{fig:gloc}
\end{center} 
\end{figure}

At this stage, we still need to obtain the surface density of the \stf rate.  
We therefore divide $\Sigst^{loc}(s,t)$ by the time-span $t$ since star formation onset so as to convert the local surface density in YSOs, $\Sigst^{loc}(s,t)$,  into the local surface density of the \stf rate, $\Sigsfr^{loc}(s,t)$: 
\begin{equation}
\Sigsfr^{loc}(s,t) = \frac{\Sigst^{loc}(s,t)}{t}\;.
\label{eq:locsfr}
\end{equation}

We note that Eq.~\ref{eq:locsfr} does not define an instantaneous value, but a time-averaged one.  This is the \sfr surface density which would be inferred by observers based on the YSO number, YSO mean mass, \stf duration, 
and the cloud surface area where YSOs have been counted.  An estimate of the instantaneous \stf rate surface density can be obtained based on the number of protostars and duration of the Class~0+Class~I phases \citep{hei10, lad13}.

How $\Sigg^{loc} (s,t)$ and $\Sigsfr^{loc}(s,t)$ relate to each other is shown in Fig.~\ref{fig:gloc} as the (blue) line with plain squares.  The model \sfing region is a clump with a radius of 6\,pc and a total mass of $10^4\,\Ms$, thus a mean volume density of $12\,\Msppp$.  This is about 60 times less dense than the median HCN 1-0 clump of \citet{wu10}.
This low-density \sfing region is representative of the highest concentration of YSOs in the MonR2 molecular cloud, one of the nearby clouds surveyed by \citet{gut11}. $t=2$\,Myr is adopted and the initial gas density profile is an isothermal sphere.  
The top and bottom panels of Fig.~\ref{fig:gloc} are identical to those of Fig.~\ref{fig:glob} except that the open symbols (time evolution of high-density clumps) have been removed for the sake of clarity.  Figure \ref{fig:gloc} shows clearly that the global and local \stf relations have different slopes.   The local relation has a slope of about two \citep[$N' \simeq 2$; ][]{gut11, par13}, which is steeper than the global \stf relations given by Equations~\ref{eq:sigsfr} ($N=1$) and \ref{eq:sigsfr_R} ($N=1.5$).  

\subsection{Global vs. Local}
\label{ssec:gl}

We can now understand the issues set in the introduction.  That is, why the observations of \citet{wu10} and \citet{hei10} have different slopes, and why, at a given gas surface density, they do not necessarily provide the same \stf rate surface density.  This may simply be the result of the intrinsic differences between the global and local \stf relations, with the former represented by the dense-clump data of \citet{wu10}, and the latter by the nearby-cloud data of \citet{hei10}.  As already stated by \citet{hei10}, this is not necessarily the result of star formation depending on the presence of massive stars.  We also note that, at this stage, this does not necessarily call for a change in the calibration between the \stf rate and infrared luminosity of molecular clumps (Eq.~\ref{eq:ken98}).  Even for a given physics of star formation -- e.g. volume-density driven star formation with a given star formation efficiency per \fft --, different slopes and different normalizations are expected when measuring the surface densities locally or globally (see Fig.~\ref{fig:gloc}).  For instance, when $\eff = 0.005$ (top panel of Figure \ref{fig:gloc}), the global \stf relation at constant mean volume density (green line) predicts a \sfr surface density five times smaller than its local counterpart (blue line with plain squares) at $\Sigg \simeq 5000\Mspp$.   
It is therefore worth emphasizing that fitting locally-obtained data with a global model could result in the erroneous conclusion that the \sfe per \fft is an increasing function of the gas surface density.

Our result that the local and global \stf relations have different slopes is reminiscent of that of \citet{lad13}.  They note that nearby molecular clouds present a power-law local \stf relation (with a slope of 2), but no global \stf relation given that their mean surface density is about constant, at $\Sigma_{gas} \simeq 40\,\Mspp$.  \citet{lad13} therefore argue against the classical Schmidt-Kennicutt law (i.e. Eq.~\ref{eq:sfla}), which extends down to $\simeq  1\,\Mspp$, representing a physical relation between the \sfr and a physical property of the cloud gas.  Instead, it results from the unresolved measurements of giant molecular clouds in the disks of galaxies, and from the corresponding beam dilution of the CO observations \citep[see][for a discussion]{lad13}.

\section{On the slope of the local star formation relation}
\label{sec:steep}
 
\citet{par13} model the relation between the local surface densities in gas and YSOs by applying a constant \sfe per free-fall time to a centrally-concentrated gas sphere.  They find a slope steeper than unity, which reflects the fact that the volume density profile of the forming cluster is steeper than that of the gas (their Figs.~1 and 3).  The model slope is $N' \simeq 2$, in good agreement with the averaged \stf relation found by \citet{gut11} for a sample of molecular clouds of the Solar neighborhood.  We remind it here for the sake of clarity: 
\begin{equation}
\frac{\Sigst}{1\Ms \cdot pc^{-2}} \simeq 10^{-3} \frac{\Sigg^2}{1\Ms \cdot pc^{-2}}\;.
\label{eq:gutsfr}
\end{equation}

\begin{figure}
\begin{center}
\epsscale{1.1}  \plotone{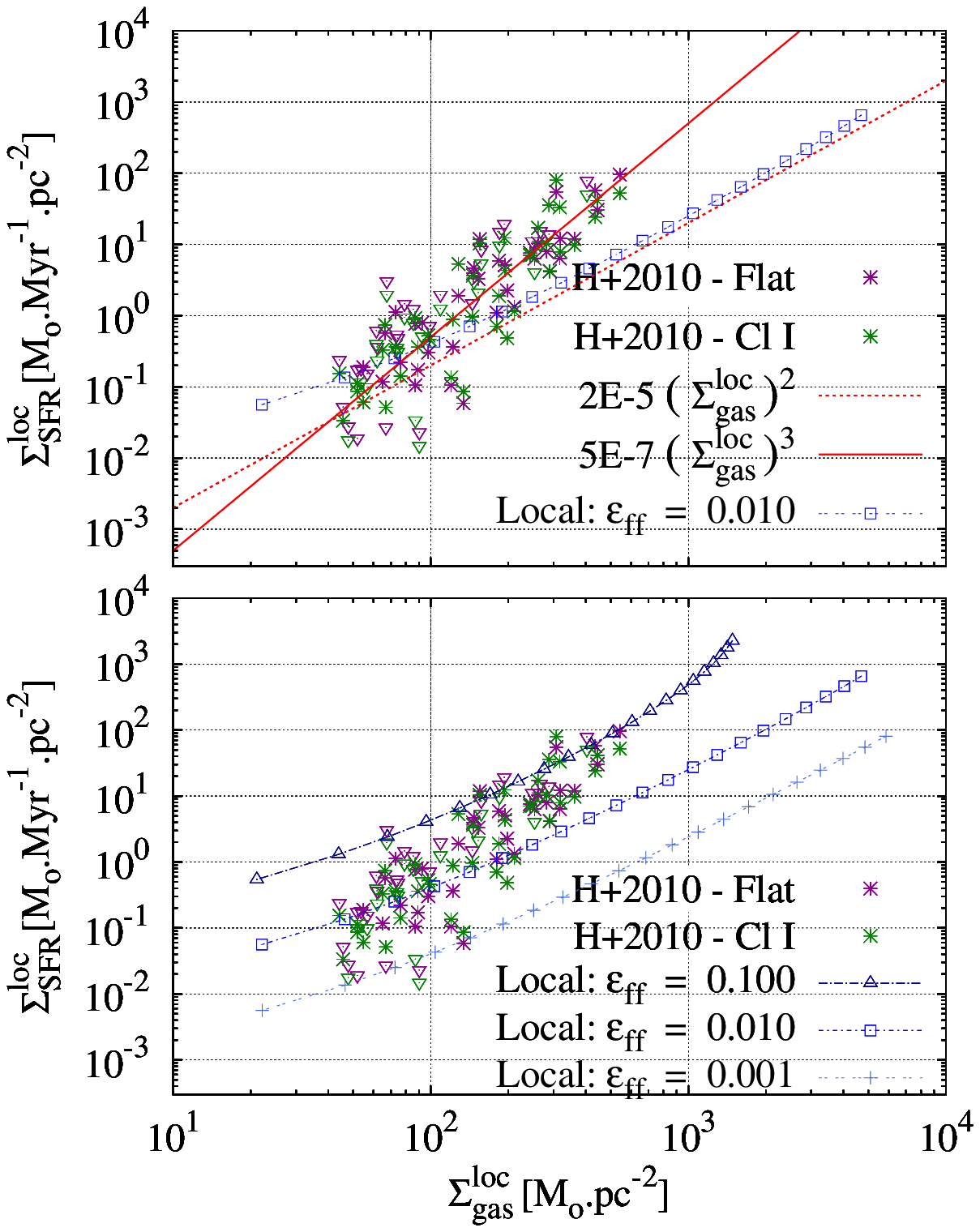}
\caption{Top panel: The asterisks and inverted triangles represent the observed local \stf relation of \citet{hei10} (with symbol- and color-codings similar to their Fig.~9; green and indigo symbols depict Class~I and Flat SED YSOs).  The two (red) symbol-free straightlines have slopes of two and three, respectively.  The (blue) line with open squares shows the model \stf relation for a constant \sfe per free-fall time ($\eff = 0.01$) applied to a sphere of gas with a mass of $10^4\Ms$, a radius of 6\,pc and a power-law initial gas density profile with a slope of $-2$.  While the data follow a slope of about three, the model with a constant $\eff$ has a slope of two.  Bottom panel: data of \citet{hei10} shown along with three models, each with its own \sfe per free-fall time, from $\eff = 10^{-3}$ (bottom line with plus-signs) up to $\eff = 10^{-1}$ (top line with open triangles).   }
\label{fig:hei10}
\end{center} 
\end{figure}

The data of \citet{hei10}, shown in Figure~\ref{fig:hei10} (color- and symbol-codings similar to their Figure~9), reveal a different picture, however.  The top panel shows that these data follow a local \stf relation with a slope of $N' \simeq 3$, rather than $N' \simeq 2$.  That is, the observations are steeper than the prediction for a constant \sfe per free-fall time (shown as the blue line with open squares).
We point out that, in Figure \ref{fig:hei10}, both the observations and the model describe a {\it local} \stf relation.  Therefore, the mismatch between both slopes cannot be ascribed to an inconsistent comparison between a global model and local observations, or vice-versa.  

Here, we speculate about two possibilities which may explain this slope difference. 
  
{\it (i)} The observations of \citet{hei10} may suggest that the \sfe per \fft increases with the gas density.  This is illustrated in the bottom panel of Figure \ref{fig:hei10}, where three models with distinct \stf efficiencies per \fft are shown along with the observations. 

An increase with gas density of the \sfe per \fft would hasten \stf in the clump inner regions.  In turn, this would yield a \stf relation steeper than $N' \simeq 2$, although it remains to be seen whether this effect is strong enough to explain the observations of \citet{hei10}.  Additional consequences would include density profiles of forming clusters steeper than what is predicted by the model of \citet{par13} (see their figs~1 and 2) and, presumably, a greater ability of their inner regions to survive the expulsion of the residual \stf gas.  We stress, however, that a proper calibration of the \sfe per \fft requires additional parameters such as the mass and volume density of the \sfing region (the cloud mass and radius adopted here are $10^4\,M_{\odot}$ and 6\,pc, as in Section~\ref{ssec:loc}).  In addition, the data of \citet{hei10}  describe entire molecular clouds, i.e. complexes of cluster-forming clumps.  Ideally, one would like to have the data as per molecular clump and to match them with models of the same mass and radius.    Therefore, the simple exercise of the bottom panel of Fig.~\ref{fig:hei10} alone does not allow one to quantify the range of $\eff$ at work in nearby molecular clouds, assuming there is one.  
It should also be noted that the \stf relation inferred by \citet{gut11} presents a range of slopes, from 1.9 (Ophiuchus molecular cloud) to 2.7 (MonR2 molecular cloud), with Eq.~\ref{eq:gutsfr} providing only a mean trend for their whole cloud sample.  The slope and universality of the local \stf relation for molecular clouds of the \SoN remain therefore debated.     \\

{\it (ii)} \citet{lom13} derive, based on a likelihood analysis, a slope of $N' \simeq 2$ for the Orion-A molecular cloud (note that their $\beta$-parameter is our $N'$ slope).  This is in agreement with the model prediction for a constant \sfe per free-fall time.  They also generate a synthetic population of protostars obeying a given \stf relation, and investigate how well it is recovered, using either a likelihood analysis or a simple least-squares fit of binned data.  Interestingly, the slope they recover with the least-squares fit is significantly steeper than the intrinsic one.  For an  input/intrinsic slope of $\beta = 1.8$, the recovered one is steeper by almost one dex, i.e. $\beta \simeq 2.7$.  A maximum-likelihood analysis is needed to recover the input slope accurately.  They ascribe the poor accuracy of the standard fit to its inability to account for the Poisson statistics of the bins, and to the loss of information inherent to data binning.  Given that the data of \citet{hei10} correspond to bins in visual extinction, hence gas surface density, it would be interesting to see whether there is a statistical bias responsible for their fairly steep slope of $N' \simeq 3$.  Such an investigation is beyond the scope of this contribution, however.     \\

An observed local \stf relation steeper than $N' \simeq 2$ can therefore result from either a statistical bias, or variations of the \sfe per \fft with the gas density (or from any other so far overlooked parameter).  This example demonstrates the need and importance of a reliable processing of the data before interpreting them physically. 

\section{From global to local: On the interpretation of the slope change}
\label{sec:break}

\subsection{A composite star formation relation}
\label{ssec:comp}

\begin{figure}
\begin{center}
\epsscale{1.1}  \plotone{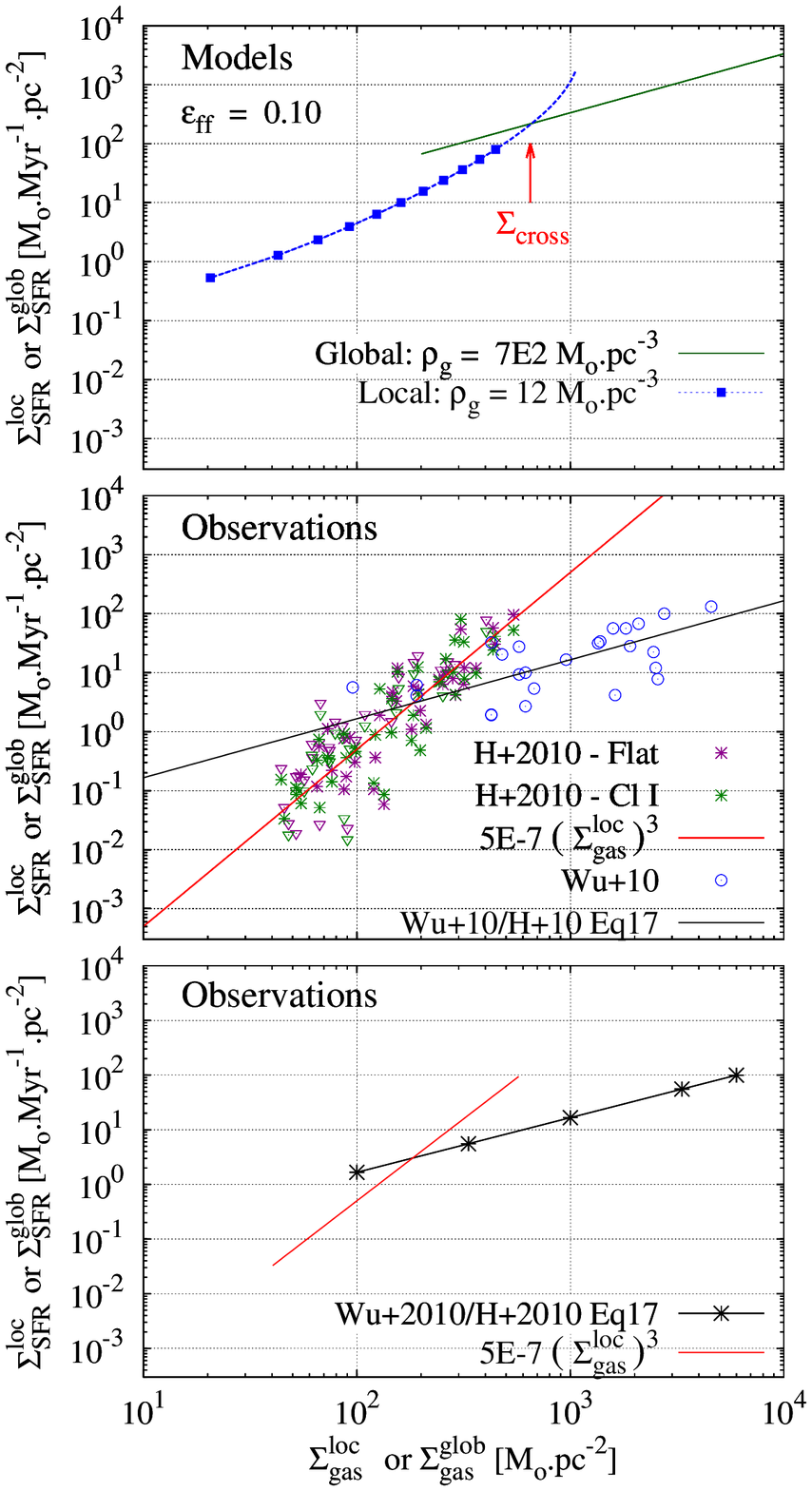}
\caption{Top panel: Model local and global relations for the mean volume densities quoted in the key.  The vertical arrow indicates the gas surface density, $\Sigma_{cross}$, at which the two \stf relations cross each other.  Middle panel: data of \citet{hei10} for a sample of nearby molecular clouds along with the data of \citet{wu10} for a sample of distant molecular clumps.  The blue open circles and the black line define a global \stf relation, while the asterisks/inverted triangles and the red line define a local relation.  Bottom panel: locus of the two data sets are shown so as to highlight that the change of slope in the \stf relation around $\Sigg \simeq 200\,\Mspp$ stems from swapping a local quantification of star formation for a global one.  }
\label{fig:hei10wu05}
\end{center} 
\end{figure}

That the local \stf relation is steeper than its global counterpart (see Fig.~\ref{fig:gloc}) can lead to the erroneous conclusion that \stf becomes less efficient at gas surface densities lower than the `break-point', that is, the point where both relations cross.  This is illustrated in the top panel of Fig.~\ref{fig:hei10wu05} where the adopted models are as in the bottom panel of Fig.~\ref{fig:gloc}, that is, a low-density clump (mean volume density of 12\,$\Msppp$) 
for the local relation, and an ensemble of high-density clumps (mean volume density of 700\,$\Msppp$) for the global relation.  In both cases, the \sfe per \fft is $\eff = 0.10$.  At a gas surface density of $\Sigg = \Sigma_{cross} \simeq 600 \Mspp$, the local relation (blue line with plain squares) crosses the global one (green line).

An observed local relation is expected to present an upper limit caused by the inability of the observing facility to resolve regions of too high stellar densities.  The highest surface density of the \sfr in \citet{hei10} is $\Sigma_{SFR,lim} \simeq 100\,\Mspp \cdot Myr^{-1}$.  Model surface densities higher than that limit are represented as a (blue) symbol-free dashed line in the top panel of Fig.~\ref{fig:hei10wu05}.  As for the observed global relation, it presents a lower limit due to the scarcity of HCN 1-0 clumps whose surface density is lower than $\Sigma_{gas, lim}\simeq200\,\Mspp$.  We impose the corresponding lower limit to the global relation (solid green line).  The combination of both aspects, namely, the change of slope from the global relation to the local one, and their respective ranges in gas and star surface densities, may at first glance suggest that star formation becomes less efficient at $\Sigg < \Sigma_{cross}$.  Yet, as we have seen, this conclusion would be erroneous.
The change in slope stems from modifying how the gas and star surface densities are measured, that is, through a succession of shells in the local relation, while it is averaged over the whole clump extent in the global relation (see Fig.~\ref{fig:sk}).  
If, additionally, the observed local relation is artificially steepened, as described by \citet{lom13}, then the impression of \stf becoming less efficient towards lower gas surface densities is strengthened even more.  

The middle panel of Fig.~\ref{fig:hei10wu05} shows the data of \citet{hei10} (local relation) and of \citet{wu10} (global relation).  \footnote{As for the clumps of \citet{wu10}, only those more luminous than $L_{IR} = 10^{4.5}\,L_{\odot}$ are shown.  For such clumps, the high-mass tail of the stellar IMF is likely to be well-sampled \citep{wu05, hei10}.}  Their respective slopes are $\simeq 3$ (red line) and $\simeq 1$ (black line; Eq.~17 in \citet{hei10} reproduced as Eq.~\ref{eq:hei10} in this paper).  The locus of the observations is further sketched in the bottom panel of Fig.~\ref{fig:hei10wu05} for the sake of clarity, where the broken power-law of the composite \stf relation is clearly highlighted.      

The top (models) and bottom (locus of the observations) panels of Fig.~\ref{fig:hei10wu05} present a similar pattern, that is, an apparent decrease in \stf activity below a break-point.  This decrease is apparent only: it is driven by the swap of the global \stf relation for a local one towards lower gas surface densities.  This also explains why a double power-law yields a better fit to the combined data set of the middle panel than does a single power-law \citep[see section 3.2.1 in][]{hei10}.  We remind again that in the top panel of Fig.~\ref{fig:hei10wu05}, the underlying physics of star formation is the same in both the local and global relations.  That is, the change of slope from the global relation to the local one does not per se indicate that \stf becomes less efficient with decreasing gas surface density.         

We note that the models and observations occupy different locii in the ($\Sigg$, $\Sigsfr$) parameter space, an aspect we will discuss in section~\ref{sec:sfeff}.

\subsection{A threshold for star formation?}
\label{ssec:th}

\citet{hei10} observe in their data a steep decline in $\Sigsfr$ and $\Sigsfr/\Sigg$ at a gas surface density of $\simeq 100 - 200 \Mspp$.  They identify that steep decline as a \stf threshold in gas surface density.  While their local measurements alone may already suggest a slope change, the effect is strengthened by the combination of the local and global \stf relations all-together.  As we saw in  Section~\ref{ssec:comp}, this leads to a composite \stf relation whose physical interpretation can be spurious.  \\ 

In the disk of our Galaxy, observational evidence of a Heaviside threshold for \stf in the ($\Sigg$, $\Sigsfr$) space remains ambiguous.  \citet{gut11} detect none in their local measurements of nearby molecular clouds, consistent with the result of \citet{lad13} for the Orion~A and Taurus clouds.  However, two other clouds in the sample of \citet{lad13}, Orion~B and California, show evidence for such a threshold, the result for Orion~B being somewhat less constrained than for California \citep[see eq.~3 and table~1 in ][]{lad13}.

Although there is no clearly-established Heaviside threshold in the ($\Sigg$, $\Sigsfr$) space, \citet{lad13} find evidence for a threshold in the cumulative protostellar fraction, that is, in the evolution of the number of protostars enclosed within a given extinction level as a function of extinction.  In the case of Orion~A, the bulk of \stf is confined to the high ($\Sigma_{gas} \gtrsim 160 \Mspp$) column-density regions of the cloud.  Despite the similar values of the surface density threshold of \citet{hei10} and \citet{lad13}, it is important to underline their different physical natures.  That is, a threshold in the cumulative protostellar fraction does not equate with a threshold in the local \stf relation.  This is best understood by comparing the left and right panels of fig.~5 in \citet{lad13}.  They depict, respectively, the local \stf relation and the variation of the cumulative protostellar fraction with extinction of the Orion~A molecular cloud.  The \stf threshold identified in their right panel at $\Sigma_{gas} \simeq 160 \Mspp$ ($A_K \simeq 0.8\,mag$) does not translate into a threshold in the left panel. It should be noted that no \stf threshold is required by the model of \citet{par13}, with the exception of the minimum gas surface density required for the phase transition from an atomic to a molecular interstellar medium, i.e. $\Sigma_{gas} \simeq 9 \Mspp$ \citep{big08}.

\section{The value of the star formation efficiency per free-fall time}
\label{sec:sfeff}

\begin{figure}
\begin{center}
\epsscale{1.1}  \plotone{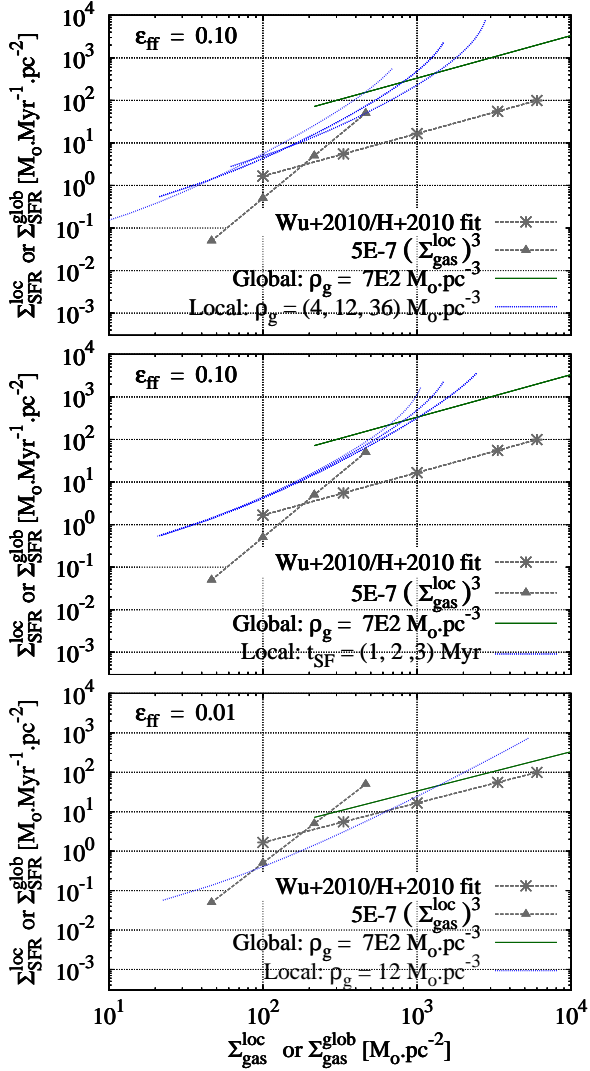}
\caption{Comparison between the locii of the data (dashed grey lines with asterisks and triangles) and model predictions (solid green line and dotted blue lines).  Top panel: variation of the mean volume density of the low-density clump, 4, 12, and 36 $\Msppp$ (dotted blue lines from left to right).  Middle panel: variation of the time-span elapsed since \stf onset, $t_{SF} = 3, 2, 1$\,Myr (blue dotted lines from left to right).  Bottom panel: the \sfe per free-fall time is decreased by an order of magnitude, i.e. $\eff = 0.01$.    }
\label{fig:pdisc}
\end{center} 
\end{figure}

While the top and bottom panels of Fig.~\ref{fig:hei10wu05} are qualitatively similar, they nevertheless present quantitative differences.  These include: a predicted slope shallower than that observed for the local \stf relation, and different normalizations for both the global and local relations.    

In Fig.~\ref{fig:pdisc}, we vary the model parameters to see whether the match between model and observations can be improved.  Dashed (grey) lines with asterisks and triangles show the locii of the diagram occupied by the observations.  The global and local models are depicted as the (solid) green and (dotted) blue lines, respectively.  In the top panel, the mean volume density of the low-density clump is varied by almost an order of magnitude, from 4 to 36\,$\Msppp$.  With a higher volume density, the local \stf relation reaches higher surface densities, but without changing its slope or normalization markedly.  In the middle panel, the local relation is presented for various time-spans since \stf onset, i.e. $t_{SF} = 1, 2$ and 3\,Myr.  A difference is noticeable in the high-density regime, which corresponds to the clump central regions where the free-fall time is shorter hence gas depletion quicker: longer time-spans yield smaller gas surface densities, as expected, but do not affect the overall trend.  In the bottom panel, we decrease the \sfe per free-fall time by an order of magnitude, i.e. $\eff = 0.01$ instead of $\eff = 0.1$ in the top and middle panels.  The models and the data now occupy similar locii in the ($\Sigg$, $\Sigsfr$) diagram, although the discrepancy in the slope of the local relation remains.  As we saw in Section~\ref{sec:steep}, steepening the model slope could be obtained by increasing the \sfe per free-fall time with the gas density.    
Another possibility is that the local relation inferred by \citet{hei10} is artificially steepened as a consequence of binning the data in visual extinction hence gas surface density \citep{lom13}.  If this is the case, their observed local relation cannot be used to estimate the \sfe per free-fall time.   

Using the mean \stf relation of \citet{gut11} (see Eq.~\ref{eq:gutsfr} in this paper), \citet{par14} estimate the \sfe per free-fall time to be between 3 and 10\,\%, depending on the adopted duration of the Class~II phase \citep[see also][]{par13}.  In contrast, the bottom panel of Fig.~\ref{fig:pdisc} shows that $\eff = 0.01$ is still slightly too large for the high-density model (solid green line) to match the location of the HCN 1-0 clumps of \citet{wu10} (dashed line with asterisks).  We indeed saw in the top panel of Fig.~\ref{fig:glob} that the high-density clump data suggest $\eff = 0.5$\,\%.  How can that value be reconciled with the significantly higher estimate of \citet{par14}?
There might be different explanations. 

\subsection{Underestimated clump infrared luminosity}
\label{ssec:LIR}
\begin{figure}
\begin{center}
\epsscale{1.1}  \plotone{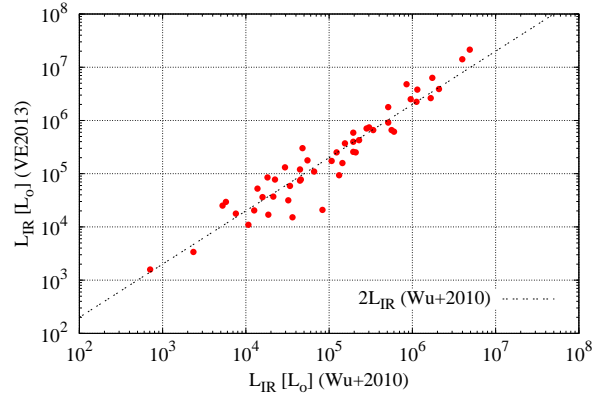}
\caption{Comparison between the infrared luminosities of dense molecular clumps as obtained by \citet{wu10} and \citet{vut13} ($x$- and $y$-axis, respectively).  The luminosities derived by \citet{vut13} are a factor of about two higher than those derived by \citet{wu10} because their adopted aperture is larger than the $IRAS$ beam size}
\label{fig:lircomp}
\end{center} 
\end{figure}

The total infrared luminosity of the clumps in \citet{wu10} is underestimated as a result of being derived from the four $IRAS$ bands \citep[see Eq.~3 in][]{wu10}.  \citet{vut13} note that most clumps are extended sources, i.e. the average source size is larger than the $IRAS$ beam.  They therefore re-estimate the infrared fluxes by performing aperture photometry of the clumps, instead of adopting the data from the $IRAS$ point-source catalog.  \citet{vut13} note that their $L_{IR}$ estimate is higher than what is obtained from the $IRAS$ point-source catalog by a factor of two on average.  We illustrate this in Fig.~\ref{fig:lircomp} which compares the infrared luminosities of dense molecular clumps as obtained by \citet{wu10} and \citet{vut13}.  Doubling the clump infrared luminosity raises the global \stf relation by a factor of two in the ($\Sigg$, $\Sigsfr$) parameter space and yields therefore twice as high an estimate of the $\eff$, i.e. $\eff = 0.01$ rather than $\eff = 0.005$.   

\subsection{Overestimated $L_{IR}$ - $SFR$ ratio}
\label{ssec:LSFR}

Although most of the stellar light is reprocessed into the infrared for both starburst galaxies and dense molecular clumps, \citet{kru07} point out that Eq.~\ref{eq:ken98} is not appropriate to recover the \sfr of individual clumps from their infrared luminosity.  This issue is also discussed by \citet{hei10} who quote that the $L_{IR} - SFR$ calibration of \citet{ken98b} is likely to underestimate the \sfr of an individual molecular clump.  
This is because during the first few Myr of star formation inside a clump, the ratio between its infrared luminosity and \sfr is lower than given by Eq.~\ref{eq:ken98}, even when the stellar initial mass function is evenly populated up to its upper mass limit.  Figure~1 in \citet{kru07} shows that
the ratio between the luminosity and \sfr of a stellar population is lower than the \citet{ken98b} relation by a factor of 5 at $t_{SF} \simeq 1$\,Myr and by a factor of 10 at $t_{SF} \simeq 0.5$\,Myr (see also \citet{urb10} and section~3.2 of \citet{hei10} for a discussion).  
The situation is even worse for forming-clusters not massive enough to sample the full mass range of the stellar IMF.  
In the case of the Orion~A and Orion~B clouds, \citet{lad12} show that the \sfr obtained from the FIR-luminosity underestimates that obtained from direct-YSO counting by a factor of 8.  

These two factors -- time-evolving infrared luminosity and under-sampling of the stellar IMF -- therefore yield luminosity-to-\sfr ratios significantly lower than what is given by Eq.~\ref{eq:ken98}.  
Thus the observed \stf relation of dense clumps (grey line with asterisks in Fig.~\ref{fig:pdisc}), obtained from Eq.~\ref{eq:ken98}, may underestimate the actual \sfr surface density.  Accordingly, the clump $\eff$ may be higher than the 1\,\%-estimate inferred in Section~\ref{ssec:LIR}.  That is, if Eq.~\ref{eq:ken98} is higher than the actual luminosity-to-\sfr ratio by a factor of 5, the clump \sfr derived from Eq.~\ref{eq:ken98} will be underestimated by a factor of 5, and so will be the estimated \sfe per free-fall time.  Therefore, the dense-clump data of \citet{wu10} do not exclude $\eff$ estimates of a few percent.  We also note that the majority of the clumps studied by \citet{wu10} have masses smaller than 3000\,$\Ms$ (their fig.~30).  They will therefore produce stellar populations where the high-mass tail of the stellar IMF is not, or poorly, sampled \citep[although a fraction of star formation may occur at densities lower than $\simeq 700\,\Msppp$, i.e. beyond the HCN $1-0$ half-peak contour; see figs.~1 and 7 in][]{par13}.  The small cluster mass would strengthen the discrepancy between Eq.~\ref{eq:ken98} and the actual luminosity-to-\stf rate ratio of the clumps, leading to even greater upward corrections of their \sfe per free-fall time.     \\

And there is yet another source of uncertainty in deriving the \sfr of a clump from its infrared luminosity.  This additional uncertainty is related to the clump star formation history.  Figure~1 in \citet{kru07} builds on a constant rate of star formation.  Yet, indications of a star formation activity declining with time have been obtained for some star-forming regions, e.g. Chamaeleon~I \citep{luh07, bel11}.  For molecular clumps in near-equilibrium, a decrease with time of their star formation rate is expected as a result of gas depletion as stars form, and of the corresponding increase of the clump free-fall time \citep[see][for an in-depth discussion]{par14}.  An additional requirement to convert the clump infrared luminosity into their \sfr is therefore a model accounting for the actual star formation history, which may not be one of a constant \stf rate.  

Finally, we note that, even on a galaxy scale, \citet{cp11} recommend an upward revision of extragalactic \stf rates.

\subsection{Overestimated local \stf relation?}
\label{ssec:sfl}
It may also be that \citet{par13} overestimated the \sfe per \fft because the \stf relation of \citet{gut11} is itself overestimated.   

\citet{lom13} find that, for the Orion-A molecular cloud, the surface density of protostars (in number counts) is:     
\begin{equation}
\Sigma _{proto} \simeq 1.65 \left( \frac{A_K}{mag} \right)^{2.03} [{\rm stars} \cdot pc^{-2}]\,,
\label{eq:lomsfln}
\end{equation}
with $A_K$ the extinction in the $K$-band.  
Adopting a mean YSO mass of $0.5\Ms$ and the conversion $\frac{\Sigg}{\Mspp} = 197 \frac{A_k}{mag}$ gives (in mass counts):
\begin{equation}
\Sigst = 2 \cdot 10^{-5} \Sigg^{2.03}\,,
\label{eq:lomsflm}
\end{equation}
with $\Sigst$ and $\Sigg$ in units of $\Mspp$.

This is a factor of 50 less than found by \citet{gut11}.  It is important to realize, however, that the \stf relation of \citet{gut11} covers Class~I and Class~II objects, while \citet{lom13} include only Class~I protostars.  Given that in nearby \sfing regions the total number of Class~I and Class~II objects is higher than the number of Class~I objects by a factor of 3-to-10 \citep[see Table~1 in][]{gut11}, the actual discrepancy between Eq.~\ref{eq:gutsfr} and Eq.~\ref{eq:lomsflm} is probably no higher than a factor of 10.  One might speculate that the lower normalization found by \citet{lom13} is conducive to a $\eff$-estimate lower than the 10\,\% found by \citet{par13} \citep[see also][for a detailed discussion of the Orion~A and B data, and of the role played by the resolution of the extinction maps]{lad13}.  \\

In summary, there are various possibilities to bridge the gap between the $\eff$ as estimated for individual molecular clumps ($\eff \simeq 0.5$\,\%, top panel of Fig.~\ref{fig:glob}) and for star-forming clouds of the Solar Neighbourhood \citep[$\eff \simeq 3-10$\,\%,][]{par14}.  {\it (i)} The $\eff$ derived for molecular clumps is underestimated because of an underestimated clump infrared luminosity (Section \ref{ssec:LIR}).  {\it (ii)} The \sfr of clumps may be underestimated when using the \citet{ken98b} relation (Eq.~\ref{eq:ken98}) as this one was established for galaxies, and is not suitable for molecular clumps whose hosted cluster is not fully developed yet (Section \ref{ssec:LSFR}).  {\it (iii)} On the other hand, the $\eff$ obtained by \citet{par13} may be too high because the local \stf relation of \citet{gut11} overestimates the surface density of YSOs at a given gas surface density (Section \ref{ssec:sfl}).  A detailed comparison of the various \stf relations for nearby molecular clouds is beyond the scope of our paper, however.        \\

\section{Conclusions}
\label{sec:conclu}
The key-points to be taken from our paper are the following: \\

- The change of slope between the local and global \stf relations results from two different ways of measuring the gas and \sfr surface densities of molecular clumps.  The global relation describes how the clump {\it mean} surface densities relate to each other.  To define it observationally therefore requires a {\it population} of clumps.  In contrast, the local relation relates the gas and \sfr surface densities measured in a {\it set of nested intervals inside a given}  clump.  As such, a local relation can be obtained for one single clump already (see Fig.~\ref{fig:sk}).  Whether the gas and star surface densities are measured globally or locally leads to different \stf relations {\it even when the underlying physics of star formation is the same}.  We refer the combined local and global \stf relations as a composite \stf relation. \\

- As a result of the above definitions, the local \stf relation has so far been measured in nearby molecular clouds only, since their YSOs can be detected on a one-by-one basis.  A global \stf relation can be obtained for both nearby molecular clouds, and distant compact molecular clumps \citep[see][]{hei10, lad13}.  In this paper, we have considered the global relation of distant high-density clumps.    \\

- That the slope of the local relation for nearby molecular clouds is steeper than that of the global relation for distant compact clumps does not necessarily imply that \stf becomes less efficient with decreasing gas surface densities.  Actually, the slope of the \stf relation also depends on how measurements are performed, i.e., on a global or local scale.  A similar conclusion is reached by \citet{lad13} about giant molecular clouds: there is a local \stf relation within them, but no global \stf relation between them. \\

- Contrary to distant molecular clumps, nearby molecular clouds are devoid of massive stars.  Again, that difference is not necessarily the reason as to why the global and local \stf relations differ.  \\

- That nearby-cloud regions and distant compact clumps show different \sfr surface  densities at a given gas surface density can be explained by the different nature of the local and global relations.  It does not necessarily require the $SFR - L_{IR}$ relation to be re-calibrated.    \\

- It is crucial not to apply a global star formation relation to locally-made measurements as one would otherwise wrongly infer that the \sfe per \fft increases with gas density. \\

- A power-law index $N$ for the global \stf relation different from $1.5$ does not necessarily preclude that a fixed fraction of gas is converted into stars every free-fall time.  The slope of the global relation is driven by the mode of \stf (i.e. is the \sfe per \fft constant, or is it gas-density dependent?), {\it and} by the clump mass-radius relation.  



\acknowledgments
GP thanks an anonymous referee for a constructive and careful report.  She acknowledges support from the Olympia-Morata Program of Heidelberg University, and from the Sonderforschungsbereich SFB 881 "The Milky Way System" (subproject B2) of the German Research Foundation (DFG). She thanks Eva Grebel for a careful reading of the initial version of the manuscript.







\end{document}